\documentclass[journal]{IEEEtran}
\usepackage{blindtext}
\usepackage{graphicx}
\usepackage{amssymb}
\usepackage{amsmath}
\usepackage{amsthm}
\usepackage{setspace}
\usepackage{amsthm}
\usepackage{mathrsfs}
\usepackage{upgreek}
\usepackage{subcaption}
\usepackage{algorithm2e}
\usepackage{multirow}
\usepackage{multicol}
\usepackage{float}

\usepackage[letterpaper, margin=1in]{geometry}

\usepackage{cite}

\ifCLASSINFOpdf
\else
\fi
\begin{document}

\singlespacing

\title{Modeling Traffic Networks Using Integrated Route and Link Data}
%
%
%

\author{Xilei Zhao
        and James C. Spall
\thanks{This work has been submitted to the IEEE for possible publication. Copyright may be transferred without notice, after which this version may no longer be accessible.}
\thanks{X. Zhao is with the H. Milton Stewart School of Industrial and Systems Engineering, Georgia Institute of Technology, Atlanta, GA, 30332 USA (e-mail: xilei.zhao@isye.gatech.edu).}
\thanks{J. C. Spall is with the Johns Hopkins University, Applied Physics Laboratory, Laurel, MD 20723 USA and with the Department of Applied Mathematics and Statistics, Johns Hopkins University, Baltimore, MD 21218 USA (e-mail: james.spall@jhuapl.edu).}}

\maketitle

\begin{abstract}

Real-time navigation services, such as Google Maps and Waze, are widely used in daily life. These services provide rich data resources in real-time traffic conditions and travel time predictions; however, they have not been fully applied in transportation modeling. This paper aims to use traffic data from Google Maps and applying cutting-edge technologies in maximum likelihood estimation to model traffic networks and travel time reliability. This paper integrates Google Maps travel time data for routes and traffic condition data for links to model the complexities of traffic networks. We then formulate the Fisher information matrix and apply the asymptotic normality to obtain the probability distribution of the travel time estimates for a random route within the network of interest. We also derive the travel time reliability by considering two levels of uncertainties, i.e., the uncertainty of the route's travel time and the uncertainty of its travel time estimates. The proposed method could provide a more realistic and precise travel time reliability estimate. The methodology is applied to a small network in the downtown Baltimore area, where we propose a link data collection strategy and provide empirical evidence to show data independence by following this strategy. We also show results for maximum likelihood estimates and travel time reliability measures for different routes within the network. Furthermore, we use the historical data from a different network to validate this approach, showing our method provides a more accurate and precise estimate compared to the sample mean of the empirical data.

\end{abstract}


%
\IEEEpeerreviewmaketitle

\section{Introduction}

Popular navigation services, such as Google Maps, Waze, and Apple Maps, are widely used by drivers throughout the world to plan trips and optimally navigate real-time traffic to avoid congestion \cite{Vasserman2015}. The services use position data (or GPS data) of smartphones to obtain real-time traffic information (such as traffic conditions, car accidents, and road closure) and then optimize route calculation for individual vehicles that use the navigation software \cite{Jeske2013}. Many of the navigation services provide application program interfaces (APIs) for users to access various types of data such as distance matrices, predictive travel time, routing in traffic, real-time traffic, street view, and so on \cite{GoogleMaps}. The navigation services provide free and comprehensive real-time data resources that overcome some of the limitations of traditional data sources, such as sensors, cameras, and probe vehicles. These traditional data sources are usually hard to acquire, expensive to purchase, limited in quantity, and biased in data sampling (e.g., using taxi data to represent all drivers’ behavior). The principal objective of the paper is to overcome the limitations of traditional data sources and traditional modeling techniques through the use of statistical tools and data provided by Google Maps to model traffic networks and travel time reliability.

Modeling transportation networks has long been an important research topic in the field (e.g., \cite{Merchant1978,Daganzo1994,kotsialos2002traffic, celikoglu2007dynamic,Ben-Akiva2012,ran2012dynamic,chiou2017robust, du2018traffic}). However, previous models have many limitations, such as impractical assumptions in modeling and complicated model structure. This paper aims at partially addressing these limitations and proposing a novel approach to building models of transportation networks by applying statistical tools to integrate route and link data from Google Maps. Specifically, we model a transportation network as a multi-level system of links and selected routes. Within the network, we collect data on the traffic conditions on all links and the origin-destination (O-D) travel time for a set of specific routes. We compute maximum likelihood estimates (MLEs) for the mean output for the success rates (non-congestion: success; congestion: failure) for links, which can also be used to estimate travel times for arbitrary routes. A major reason for using the MLE-based route/link technique is that complicated connections exist between the route traffic behavior (travel times over routes) and the link traffic flow (success rates of links). That is, the complexities of network traffic (e.g., traffic incidents, work zones, bad weather, pedestrian behavior, and poor traffic signal timing) and their interactions are difficult to mathematically model, but the MLEs based on real-world data can make full use of information at both link and route levels to properly represent these connections and implicitly capture the dynamics of traffic. 

The multi-level systems method for general full systems with binary subsystems was first proposed and formulated in \cite{Spall2014}. Spall derived the asymptotic normality and confidence intervals for reliability estimates based on the MLE-based full-system/subsystem model \cite{Spall2012}. Additionally, a couple of numerical examples were applied to demonstrate the feasibility of the method \cite{Spall2013}. The method has also been applied to a single route in downtown Baltimore, which has been proved simple and easy-to-implement \cite{Zhao2016}; this paper extends the idea to full traffic networks. 

This paper generalizes the framework in \cite{Zhao2016} so that we may apply the traffic network model to generate travel time distribution and predict travel time reliability over arbitrary networks. In the context here, reliability is defined as the consistency or dependability in travel times measured from day-to-day and/or across different times of the day \cite{fha2010travel}. Travel time reliability serves as a fundamental factor in modeling and understanding people's travel behavior, representing the temporal uncertainty experienced by travelers in their movement between any O-D pairs in a network \cite{carrion2012value}. This topic has also been studied extensively in the past two decades (e.g., \cite{chen2003travel, clark2005modelling, pu2011analytic,carrion2012value,uchida2015travel}). As pointed out in \cite{pu2011analytic}, a frequent method of computing travel time reliability measures is to use empirical data directly without fitting the data to statistical distributions, but this method could be problematic by overlooking the characteristics of the underlying distributions for travel times. Moreover, our method can be applied to reliability estimation for travel demand models where travel time reliability needs to be computed for O-D matrices. Since link-level reliability measures such as standard deviation are not additive for computing O-D level reliability due to link dependence issues \cite{gupta2018incorporation}, our method presents a new approach to tackling this issue by integrating route and link data using MLE and developing a link data collection strategy to reduce the data dependence issue at the link level.

In this paper, we start from empirical data from Google Maps to build up the probability distribution of travel times by combining the uncertainties of travel times and travel time estimates. After obtaining the transportation network model using data from Google Maps, we derive the Fisher information matrix (FIM) and use it to generate the probability distribution of travel time estimates for an arbitrary route within the network. Then, based on the probability distribution of travel times, we compute different travel time reliability measures in the field (i.e., 95th percentile travel times, standard deviation, coefficient of variation, buffer index, and planning time index). Our method is capable of providing good predictions with limited data points and shows the ability of providing travel time reliability measures for any routes within the transportation network, not only the routes for which data are collected.


The remainder of the paper is organized as follows: in Sect. II, we introduce the mathematical modeling process, including the maximum likelihood (ML) formulation, route/link relationship derivation, and parameter estimation. Also, we describe the process of calculating the FIM and using the asymptotic normality result to construct probability distribution of travel time estimates. In Sect. III, we derive the travel time reliability by combining uncertainties of travel times and travel time estimates. In Sect. IV, we give a numerical example for downtown Baltimore, in which we also propose a link data collection strategy and provide empirical evidence to show independence of data by following this strategy. Additionally, we provide MLE results and the travel time reliability results for different routes in the network. We validate the MLE-based route/link technique in Sect. V. Finally, we conclude the paper by discussing the strengths and limitations of the approach, and suggest areas for future study.

\section{Methodology}

\subsection{Maximum Likelihood Estimation}
The ML formulation involves a parameter vector $\boldsymbol\uptheta$ to be estimated and a log-likelihood function $\log{L(\boldsymbol\uptheta)}$ to be maximized. The method of ML is a powerful tool for estimating parameters and is perhaps the most popular general method in practice \cite{Scholz2006}, relative to other statistical methods such as least squares or method of moments. Next, let us introduce some basic concepts and definitions of MLE. According to the definition of MLE (for example, see \cite[pp. 267--268]{Rice2006}), suppose that random variables $X_1,..., X_n$ have a joint density or frequency function $p(x_1, x_2,..., x_n|\boldsymbol\uptheta)$. Given observed values $X_i = x_i$, where $i$ = 1,..., $n$, the likelihood of $\boldsymbol\uptheta$, conditioned on $x_1, x_2,..., x_n$, is defined as
\begin{equation*}
L(\boldsymbol\uptheta)=p(x_1, x_2,..., x_n|\boldsymbol\uptheta).
\end{equation*}
The MLE of $\boldsymbol\uptheta$ maximizes the likelihood function by fully using the observed data. In the case of independent, identically distributed (i.i.d.) data, the log-likelihood function has the generic form:
\begin{equation*}
\log{L(\boldsymbol\uptheta)}=\log \prod_{i=1}^{n}{p(x_i|\boldsymbol\uptheta)}=\sum_{i=1}^{n}\log{{p(x_i|\boldsymbol\uptheta)}}.
\end{equation*}

\subsection{Basic Definitions and Assumptions for Transportation Network}

Let us introduce some basic definitions before diving into the details on how to use ML to model transportation network. As shown in Fig. 1, taking a small general transportation network as an example, we first define the boundary of the network, and in this example, the boundary is square ACIG. All the traffic links within square ACIG are considered for analysis. Each node within the network represents a unique intersection. Note that the different directions of travel in a road are considered as two distinct links. For example, in Fig. 1, link AB, from west to east, and link BA, from east to west, are treated as two separate links.

Consider a transportation network system that consists of $p$ links (subsystems). Traffic conditions on links are modeled as binary: ``0" (``failure'') for congested links and ``1" (``success'') for non-congested links. We assume that data for all the links, including within and across the links, are independent. Data are collected on different days to help ensure independence. The data for link $j$, where $j = 1, 2, \cdots, p$, are i.i.d., because we suggest collecting one data point for link $j$ at a specific time on one day; that is to say, for link $j$, data collected on Day 1 are independent of data collected on Day 2. We do not assume data across links are identically distributed; that is, the success probability generally varies by link. For data across links at a given time and day, distant links can be viewed as nearly independent, whereas the traffic conditions of adjacent links may influence each other. Therefore, we propose a novel link data collection strategy in Subsection IV-B to resolve the data problem of potential dependence in link data for links that area near one another. Note the inherent tradeoff: We want to collect as much data as possible on each day in order to rapidly build the dataset for estimation of the network parameters, yet we want to minimize the amount of data collected each day in order to help ensure statistical independence of the measurements.

\begin{figure}[ht!]
    \centering
    \includegraphics[width=0.42\textwidth]{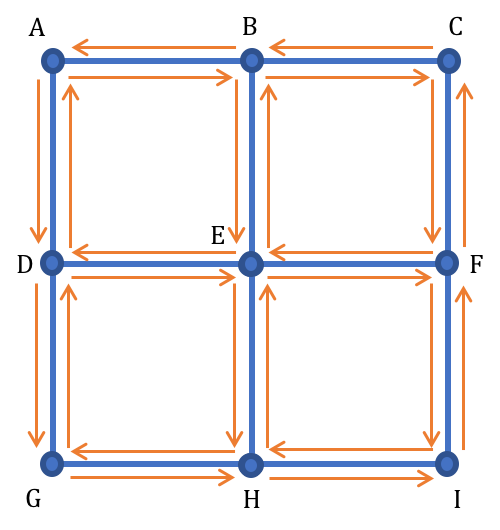}
    \caption{A general transportation network with nodes A-B-C-D-E-F-G-H-I-J.}
\end{figure}

A route (full system) is defined as the travel time from origin to destination through a specific path. We assume the route outputs (the O-D travel times along a specific route) follow the log-normal distribution. The log-normal assumption for travel times has been applied in many previous studies (for example, \cite{ElFaouzi2007,pu2011analytic,Zhao2016}). The logic behind it is simple and straightforward. First, a log-normal distribution is defined on positive real numbers, which well fits the nature of travel time. The probability distribution function for a log-normal distribution has most of its area near the mean and median travel time, but it is skewed to the right, with the right tail representing travel times with traffic delay. In \cite{Zhao2016}, we tested the distribution of travel time data collected from Google Maps against the log-normal assumption for several routes in Baltimore, and large statistical $P$-values were obtained (well above the common rejection thresholds of 0.05 or 0.01), indicating we cannot reject the null hypothesis that travel time data are consistent with a log-normal distribution.

One route typically cannot go through all the links in the network, and, in practice, people are very unlikely to drive in circles. Therefore, we collect data for several routes in order to cover all the traffic links within the network. We assume that data for all the routes are independent. Even though route data might have some statistical dependence across different routes at a given day and time, we try to minimize the dependence by properly choosing routes to minimize shared information. Note that formal experimental design for route data collection \cite[Chap. 17]{Spall2003} might be used here for collecting data efficiently and optimally, but we do not consider that in this paper. It is also worth pointing out that route data and link data are not collected on the same day in order to ensure independence.

\subsection{Maximum Likelihood Function for Transportation Network}

Let us now define $\boldsymbol{\uptheta}$ and describe our notation for the data. We use a semicolon to represent a separate row for convenience (e.g., $[a, b; c, d]$ denotes a $2\times2$ matrix with rows $a, b$ and $c, d$). Suppose that data are collected for $r$ routes in the network. Let $\boldsymbol\Pi = \big[ \upomega_1, \upsigma_1^2; \upomega_2, \upsigma_2^2; ...;\upomega_r, \upsigma_r^2 \big]$ represent an $r$-by-2 matrix with  $\upomega_i$ and $\upsigma_i^2$ representing unknown means and variances of the normally distributed logarithm of the outputs of the $r$ routes. Let $\uprho_j$ represent the success probabilities for link $j$, $j = 1, 2, ..., p$. The parameter vector $\boldsymbol\uptheta \equiv [\uprho_1, \uprho_2, \uprho_3, ..., \uprho_p]^T$; elements in $\boldsymbol\Pi$ are not included in the parameter vector to be estimated because they are uniquely determined by $\boldsymbol\uptheta$ and relevant constraints. Let $\boldsymbol{T} = \{ T_{11}, T_{12}, ..., T_{1,k(1)}; T_{21}, T_{22}, ..., T_{2,k(2)}; ...; T_{r1},$ $T_{r2}, ..., T_{r,k(r)} \}$ indicate the collection of observed, scalar-valued travel time output $T_{q,i}$ from the data collected on day $i, i = 1, 2, ..., k(q)$ for route $q, q = 1, 2, ..., r$. Because we assume the route outputs are log-normally distributed, then we let $\boldsymbol{Z} = \{ Z_{11}, ..., Z_{1,k(1)};...; Z_{r1}, ..., Z_{r,k(r)} \} = \{ \log(T_{11}), ..., \log(T_{1,k(1)});...; \log(T_{r1}),...,\log(T_{r,k(r)}) \}$ represent the normally distributed collection of log-transformed route outputs, which can facilitate the following derivation. 

We now derive the log-likelihood function based on the full set of link and route data. According to the definition of log-normal distribution and the properties of independent data, the log-likelihood function for route outputs is
\begin{equation}
\sum_{q=1}^{r}\Big[-\frac{k(q)}{2}\log(\upsigma_q^2)-\frac{1}{2\upsigma_q^2}\sum_{j=1}^{k(q)}(Z_{qj}-\upomega_q)^2 \Big] + \text{constant},
\end{equation}
where the route parameters (i.e., $\upomega_q, \upsigma_q, q = 1, 2, ..., r$) can be fully represented as functions of $\boldsymbol\uptheta$ (which will be discussed later). Let $X_{ji}$ represent the $i$th output of the $j$th link, indicating traffic conditions (failure ``0" or success ``1") on the link $j$. Thus, the number of successes in $n(j)$ i.i.d. data on link $j, j = 1, 2,..., p$, can be expressed as
\begin{equation*}
S_j \equiv \sum_{i=1}^{n(j)}X_{ji}.
\end{equation*}
Because the $X_{ji}$ follow a Bernoulli distribution, the log-likelihood function of link outputs is
\begin{equation}
\sum_{j=1}^{p}\big[S_j\log(\uprho_j)+(n(j)-S_j)\log(1-\uprho_j)\big].
\end{equation}
By adding (1) and (2), the log-likelihood function for the entire system, including all the route data and the link data, is:
\begin{small}
\begin{align}
\log{L(\boldsymbol\uptheta)}=\sum_{q=1}^{r}\Big[-\frac{k(q)}{2}\log(\upsigma_q^2)-\frac{1}{2\upsigma_q^2}\sum_{j=1}^{k(q)}(Z_{qj}-\upomega_q)^2 \Big] \nonumber \\ 
+\sum_{j=1}^{p}\big[S_j\log(\uprho_j)+(n(j)-S_j)\log(1-\uprho_j)\big] + \text{constant}.
\end{align}
\end{small}

Let $[\upomega_q,\uptheta_q^2]^T \equiv [h_{q1}(\boldsymbol{\uptheta}), h_{q2}(\boldsymbol{\uptheta})]^T $ represent the relationship between the parameters of the log-normal distribution of route $q$ and the $\uprho$ links. In order to maximize Eqn. (3), we differentiate the log-likelihood function to obtain the score vector:
\begin{multline} 
\frac{\partial\log{L(\boldsymbol\uptheta)}}{\partial\boldsymbol\uptheta}
= \sum_{q=1}^{r} \Bigg[ -\frac{k(q)}{2\upsigma_q^2}\boldsymbol{h}_{q2}'(\boldsymbol\uptheta) + \frac{\boldsymbol{h}_{q2}'(\boldsymbol\uptheta)}{2\upsigma_q^4}\\
\times \sum_{j=1}^{k(q)}(Z_{qj} - \upomega_q)^2
+ \frac{1}{\upsigma_q^2}\boldsymbol{h}_{q1}'(\boldsymbol\uptheta)\sum_{j=1}^{k(q)}(Z_{qj}-\upomega_q) \Bigg]\\ 
+ \left (
\begin{array}{c}
\frac{S_1}{\uprho_1} - \frac{n_1-S_1}{1-\uprho_1}\\
\vdots\\
\frac{S_p}{\uprho_p} - \frac{n_p-S_p}{1-\uprho_p}\\
\end{array}
\right ),
\end{multline}
where $\boldsymbol{h}_{q1}'(\boldsymbol\uptheta)$ and $\boldsymbol{h}_{q2}'(\boldsymbol\uptheta)$ represent the gradient vectors of $h_{q1}(\boldsymbol\uptheta)$ and $h_{q2}(\boldsymbol\uptheta)$ with respect to $\boldsymbol\uptheta$ for $q = 1, 2,..., r$. The vector $[h_{q1}(\boldsymbol\uptheta), h_{q2}(\boldsymbol\uptheta)]^T$ relates $\boldsymbol\uptheta$ to $[\upomega_q,\upsigma_q^2]$. Next, we will show how to derive $h_{q1}(\boldsymbol\uptheta)$ and $h_{q2}(\boldsymbol\uptheta)$.

\subsection{Relationship Between Routes and Links}

Following the precedent in \cite{Zhao2016}, the typical travel time of each link under different traffic conditions (``0" or ``1") is computed as follows according to the color scheme of Google Maps:
\begin{equation}
\begin{cases}
l_j/v & \text{if the link is blue or yellow (``1"),}\\
l_j/v'& \text{if the link is red or dark red (``0"),}
\end{cases}
\end{equation}
where $v$ and $v'$ are the mean travel speeds in different traffic conditions estimated from historical Google Maps data (different from data $\boldsymbol{T}$ and $\boldsymbol{Z}$) and $l_j$ represents the length of the link $j$. We treat $v$ and $v'$ as fixed parameters, not estimates, in the analysis below. Let $\bar{X}_{j} = S_j / n(j)$ represent the observed success rate on link $j$ using only link data, for $j = 1, 2,..., p$. Based on (5), we derive the measured typical travel time on each link, say $t_j$, and its expectation as follows:
\begin{equation*}
t_j = \bar{X}_{j}\times\frac{l_j}{v}+(1-\bar{X}_{j})\times\frac{l_j}{v'},
\end{equation*}
\begin{equation}
E(t_j) = \uprho_j\times\frac{l_j}{v}+(1-\uprho_j)\times\frac{l_j}{v'}=\frac{l_j}{v'}-\frac{v-v'}{vv'}l_j\uprho_j.
\end{equation}
Then, as shown in \cite{Zhao2016}, we are able to derive the relationship between routes and links, relating $\boldsymbol\uptheta$ to  $\boldsymbol\Pi$. Specifically, for route $q$, suppose there are $m(q)$ links within this specific route, and the corresponding parameters of the links in route $q$ can be represented as a sub-sequence of $\boldsymbol \uptheta$ with $m(q)$ components. Obviously, $m(q) \leqslant p$ (the number of links within the network). For example, suppose route 1 contains link 1, 3, and 6; then, $m(1) = 3, \uprho_{1_1} = \uprho_1, \uprho_{1_2} = \uprho_3$, and $\uprho_{1_3} = \uprho_6$.

To derive the relationship between routes and links, we use two equivalent ways (one from the route perspective, the other from the link perspective) to represent the expectation and variance for the $i$th observation of the $q$th route. We then equate these results to obtain $h_{q1}$ and $h_{q2}$. Based on the log-normal assumption for the route output, we can write down the expectation and variance for $T_{qi}$ using the log-normal properties \cite[p. 212]{Johnson1994}:
\begin{equation}
E(T_{qi}) = \exp{(\upomega_q + \frac{1}{2}\upsigma_q^2)},
\end{equation}
\begin{equation}
\text{Var}(T_{qi}) = [\exp{(\upsigma_q^2)} - 1]\exp{(2\upomega_q + \upsigma_q^2)}.
\end{equation}
We then derive the expectation and variance for $T_{qi}$ by using the link information \cite{Zhao2016}:
\begin{equation}
E(T_{qi}) = \sum_{j=1}^{m(q)}\big[\frac{l_{q_j}}{v'}-\frac{v-v'}{vv'}l_{q_j}\uprho_{q_j}\big],
\end{equation}
\begin{equation}
\text{Var}(T_{qi}) = \sum_{j=1}^{m(q)}\big[ (\frac{v-v'}{vv'}l_{q_j})^2\uprho_{q_j}(1-\uprho_{q_j})\big],
\end{equation}
where $l_{q_j}$ represents the length of the link $q_j$. Note that, because the sum of link travel times along a route equals the route travel time, the travel times through intersections are considered as part of the travel times for links.

The expectation and variance derived from the routes are required to be equal to those derived from the links. That is, Eqn. (7) and Eqn. (9) are equivalent and Eqn. (8) and Eqn. (10) are equivalent. Therefore, we have developed the relationship between routes and links. Then, let $\boldsymbol{\hat\uptheta}$ represent the MLE of $\boldsymbol{\uptheta}$, $\hat\upomega_q = h_{q1}(\boldsymbol{\hat\uptheta})$ indicate the MLE of $\upomega_q$, and $\hat\upsigma_q^2 = h_{q2}(\boldsymbol{\hat\uptheta})$ indicate the MLE of $\upsigma_q^2$ (we are using the invariance property of MLE: a function of an MLE is also an MLE). Let $M_q = \sum_{j=1}^{m(q)}\big[{l_{q_j}}/{v'}-l_{q_j}\uprho_{q_j}{(v-v')}/{(vv')} \big]$; $V_q = \sum_{j=1}^{m(q)}\big\{ [l_{q_j}{(v-v')}/{(vv')}]^2\uprho_{q_j}(1-\uprho_{q_j})\big\}$. Then, we obtain $h_{q1}(\boldsymbol\uptheta)$ and $h_{q2}(\boldsymbol\uptheta)$ as
\begin{equation*}
h_{q1}(\boldsymbol\uptheta) = \log{M_q} - \frac{1}{2}\log{\left(\frac{V_q}{M_q^2} + 1 \right)},
\end{equation*}
\begin{equation*}
h_{q2}(\boldsymbol\uptheta) = \log{\left(\frac{V_q}{M_q^2} + 1 \right)}.
\end{equation*}

After obtaining $h_{q1}(\boldsymbol\uptheta)$ and $h_{q2}(\boldsymbol\uptheta)$, we are able to compute $\boldsymbol{h}_{q1}'(\boldsymbol\uptheta)$ and $\boldsymbol{h}_{q2}'(\boldsymbol\uptheta)$ in the score vector, Eqn. (4). Solving the score equation, $\partial\log{L(\boldsymbol\uptheta)/\partial\boldsymbol\uptheta} = \boldsymbol0$, yields a candidate MLE for $\boldsymbol{\uptheta}$ that reflects a careful balancing of information between the route and links. In general, the solution to the score equation is not unique and can only be achieved numerically. 

\section{Modeling Travel Time Reliability with Two-Level Uncertainties}

After modeling the transportation network by integrating route and link data, we apply the network model to generate the travel time probability distribution by taking into consideration two levels of randomness: one due to the inherent variability of traffic flow (the above-mentioned log-normal distribution) and the other due to the estimation uncertainty in $\boldsymbol{\hat \uptheta}$. We then apply the probability distribution to calculate various travel time reliability measures associated with predicted travel times for arbitrary routes in the network.

\subsection{Fisher Information Matrix Formulation and Asymptotic Normality}

Aside from determining an MLE of $\boldsymbol{\uptheta}$ (and derived parameters $\upomega_q, \upsigma_q^2$), we are also able to produce uncertainty bounds (confidence regions) on the estimates. The confidence regions are based on asymptotic normality of the estimator with a covariance matrix derived from the FIM for $\boldsymbol{\uptheta}$ \cite{Spall2014}. The FIM contains a summary of the amount of information in the data with respect to the quantities of interest (see \cite[Sect. 13.3]{Spall2003}). The FIM has multiple applications in general problems, including confidence region construction, model selection, and experimental design. In this paper, our interest centers on the use of FIM for constructing confidence regions and related quantities for the estimates, $\hat{\upomega}_q, \hat{\upsigma}_q^2$. The $p \times p$ FIM $\boldsymbol{F}(\boldsymbol\uptheta)$ for a twice-differentiable log-likelihood function, $\log L(\boldsymbol\uptheta)$ is defined as
\begin{equation*}
\begin{split}
        \boldsymbol{F}(\boldsymbol\uptheta) & = E\Bigg( \frac{\partial{\log L(\boldsymbol\uptheta)}}{\partial{\boldsymbol\uptheta}} \cdot \frac{\partial{\log L(\boldsymbol\uptheta)}}{\partial{\boldsymbol\uptheta}^T} \Bigg) \\
        & = -E \Bigg(\frac{\log \partial^2{L(\boldsymbol \uptheta)}}{\partial \boldsymbol \uptheta \partial \boldsymbol \uptheta^T} \Bigg).
\end{split}
\end{equation*}
In this paper, $\boldsymbol{F}(\boldsymbol\uptheta)$ is given by
\begin{multline}
\boldsymbol{F}(\boldsymbol\uptheta) = \sum_{q = 1}^{r} \Bigg[\frac{k(q)}{2(h_{q2}(\boldsymbol\uptheta))^2}\boldsymbol{h}_{q2}'(\boldsymbol\uptheta)\boldsymbol{h}_{q2}'(\boldsymbol\uptheta)^T\\
+ \frac{k(q)}{h_{q2}(\boldsymbol\uptheta)}
\boldsymbol{h}_{q1}'(\boldsymbol\uptheta)\boldsymbol{h}_{q1}'(\boldsymbol\uptheta)^T \Bigg] + \boldsymbol{J}(\boldsymbol\uptheta),
\end{multline}
where 
\begin{equation*}
    \boldsymbol{J}(\boldsymbol\uptheta) = \text{diag} \Bigg[\frac{n(1)}{\uprho_1(1 - \uprho_1 )},..., \frac{n(p)}{\uprho_p(1-\uprho_p)} \Bigg].
\end{equation*}
One of the most significant properties of the MLE and FIM is asymptotic normality of the estimate. Based on asymptotic distribution theory described in Spall (2014), we have (approximately)
\begin{equation}
    \hat{\boldsymbol{\uptheta}} \sim N(\boldsymbol{\uptheta}^*,  \boldsymbol{F}(\boldsymbol{\uptheta}^*)^{-1}),
\end{equation}
where $\boldsymbol{\uptheta}^*$ represents the true value of the unknown parameter vector $\boldsymbol{\uptheta}$. 

In the following derivation, $h_1(\boldsymbol{\uptheta})$ and $h_2(\boldsymbol{\uptheta})$ and associated quantities are for an $\textit{arbitrary route}$ and we are suppressing the required subscript $q$ for notational convenience. Specifically, for an arbitrary route within the network of interest (even those for which route data were not collected) and sufficiently large sample sizes, we can formulate the corresponding functions $h_1(\boldsymbol{\uptheta})$ and $h_2(\boldsymbol{\uptheta})$ for the arbitrary route, and obtain their MLEs. By the invariance of ML, we know that $h_1(\hat{\boldsymbol{\uptheta}})$ is an MLE of $h_1(\boldsymbol{\uptheta})$ and $h_2(\hat{\boldsymbol{\uptheta}})$ is an MLE of $h_2(\boldsymbol{\uptheta})$. Hence, the probability distribution for the two-dimensional vector $\boldsymbol{h}(\hat{\boldsymbol{\uptheta}}) = [h_1(\hat{\boldsymbol{\uptheta}}), h_2(\hat{\boldsymbol{\uptheta}})]^T$ as
\begin{equation}
    \boldsymbol{h}(\hat{\boldsymbol{\uptheta}}) \sim N(\boldsymbol{h}(\boldsymbol{\uptheta}^*), \boldsymbol{\Sigma}),
\end{equation}
where 2-by-2 matrix $\boldsymbol{\Sigma} = \boldsymbol{h}'(\boldsymbol{\uptheta}^*)^T \boldsymbol{F}(\boldsymbol{\uptheta}^*)^{-1} \boldsymbol{h}'(\boldsymbol{\uptheta}^*) \equiv $
$[\upsigma_{1}^2, \upsigma_{12}; \upsigma_{21}, \upsigma_{2}^2]$ and $p$-by-2 matrix $\boldsymbol{h}'(\boldsymbol{\uptheta})= [\boldsymbol{h}'_1(\boldsymbol{\uptheta}^*),$ $\boldsymbol{h}'_2(\boldsymbol{\uptheta}^*)]$. In practice, we often set $\boldsymbol{\uptheta}^*$ equal to $\boldsymbol{\hat\uptheta}_{\text{MLE}}$ (i.e., MLE value for $\boldsymbol{\uptheta}$) on the right hand side of Expressions (12) and (13). Expression (13) can be used to compute the asymptotically based uncertainty bounds for the estimated travel time of the route.

\subsection{Modeling Travel Time Reliability}

We now show how the inherent variability of travel times in routes based on the log-normal assumption can be combined with the estimated uncertainty given in (13) to produce an integrated probability distribution that best represents the overall randomness in travel times for any route in the network. In particular, we consider the two-level uncertainties to measure the travel time reliability for any routes in the transportation networks. One level of the uncertainty comes from the probability distribution of the arbitrary route's log-transformed travel time $z$ conditioned on the parameters in $\boldsymbol{h}(\hat{\boldsymbol{\uptheta}})$; the other level of the certainty comes from the probability distribution of $\boldsymbol{h}(\hat{\boldsymbol{\uptheta}})$ itself. By generating the travel time distribution $p(z)$, we can not only use the features of the distribution to construct travel time reliability measures, but also provide an informative tool to assist traffic planning to improve congestion.

To be specific, we use $h_1$ and $h_2$ to represent an arbitrary $h_1(\hat{\boldsymbol{\uptheta}})$ and $h_2(\hat{\boldsymbol{\uptheta}})$, and use $h_{1}^*$ and $h_{2}^*$ to represent their corresponding true values, $h_1(\boldsymbol{\uptheta}^*)$ and $h_2(\boldsymbol{\uptheta}^*)$, for convenience. We then write the probability density function (pdf) of $z$ as  
\begin{equation}
    p(z) = \int \int p(z|h_1,h_2)  p(h_1|h_2) p(h_2) dh_1 dh_2,
\end{equation}
where the conditional probability of $h_1$ for a given value of $h_2$ follows a normal distribution with mean equal to $h = h_{1}^* + {\upsigma}_{12}{\upsigma}_{2}^{-2}(h_2-h_{2}^*)$ and variance equal to ${\upsigma}_{11,2} = {\upsigma}_{1}^2-{\upsigma}_{12}{\upsigma}_{2}^{-2}{\upsigma}_{21}$ and the integrals are over the real line. Therefore, according to Eqn. (14), we can derive a semi-analytical solution for $p(z)$ by integrating over the relevant domain for $h_1$ and $h_2$. That is, we can analytically integrate out the variable $h_1$, while ultimately using Monte Carlo methods to integrate out $h_2$. Specifically, $p(z)$ can be written as
\begin{small}
\begin{equation}
\begin{split}
    & p(z) = \int \int p(z|h_1,h_2) p(h_1|h_2) dh_1 p(h_2) dh_2 \\
    & = \int \frac{1}{(2\pi h_2)^{1/2}(2\pi {\upsigma}_{11,2})^{1/2}} \int \exp\Bigg(-\frac{1}{2h_2}(z - h_1)^2 \\
    & -\frac{1}{2{\upsigma}_{11,2}}(h_1 - h)^2 \Bigg) dh_1 p(h_2) dh_2 \\
    & = \int \frac{1}{(2\pi h_2)^{1/2}(2\pi {\upsigma}_{11,2})^{1/2}} \exp\Bigg(-\frac{z^2}{2h_2} - \frac{h^2}{2{\upsigma}_{11,2}} \Bigg) \\
    & \times \exp{\Bigg[\frac{(z{\upsigma}_{11,2} + hh_2)^2}{2h_2{\upsigma}_{11,2}({\upsigma}_{11,2}+h_2)} \Bigg]} \times \int \exp\Bigg[ -\frac{{\upsigma}_{11,2}+h_2}{2h_2{\upsigma}_{11,2}} \\
    & \times \Bigg(h_1 - \frac{z{\upsigma}_{11,2}+hh_2}{{\upsigma}_{11,2}+h_2} \Bigg)^2 \Bigg] dh_1 p(h_2) dh_2 \\
    & = \int \frac{1}{(2\pi h_2)^{1/2}(2\pi {\upsigma}_{11,2})^{1/2}} \exp\Bigg(-\frac{z^2}{2h_2} - \frac{h^2}{2{\upsigma}_{11,2}} \Bigg) \\
    & \times \exp{\Bigg[\frac{(z{\upsigma}_{11,2} + hh_2)^2}{2h_2{\upsigma}_{11,2}({\upsigma}_{11,2}+h_2)} \Bigg]} \times \sqrt{ \frac{2 \pi h_2{\upsigma}_{11,2}}{{\upsigma}_{11,2}+h_2}} p(h_2) dh_2\\
    & = \int \sqrt{\frac{1}{2\pi ({\upsigma}_{11,2} + h_2)}} \exp{\Bigg(-\frac{z^2}{2h_2} - \frac{h^2}{2{\upsigma}_{11,2}} \Bigg)} \\
    & \times \exp{\Bigg[\frac{(z{\upsigma}_{11,2} + hh_2)^2}{2h_2{\upsigma}_{11,2}({\upsigma}_{11,2}+h_2)} \Bigg]} p(h_2) dh_2\\
    & = E_{h_2}[p(z|h_2)],
\end{split}
\end{equation}
\end{small}

\noindent where $E_{h_2}[p(z|h_2)]$ is the expectation of $p(z|h_2)$ with respect to $h_2$, and
\begin{equation*}
\begin{split}
        p(z|h_2) & = \sqrt{\frac{1}{2\pi ({\upsigma}_{11,2} + h_2)}} \exp{\big(-\frac{z^2}{2h_2} - \frac{h^2}{2{\upsigma}_{11,2}} \big)} \\ 
        & \times \exp{\big[\frac{(z{\upsigma}_{11,2} + hh_2)^2}{2h_2{\upsigma}_{11,2}({\upsigma}_{11,2}+h_2)} \big]}.
\end{split}
\end{equation*}
Here, $h_2$ follows $N(h_{2}^*, [\boldsymbol{h}_{2}'(\boldsymbol{\uptheta}^*)]^T \boldsymbol{F}(\boldsymbol{\uptheta}^*)^{-1} \boldsymbol{h}_{2}'(\boldsymbol{\uptheta}^*))$, and in practice, we usually set $h_{1}^* = h_1(\boldsymbol{\hat\uptheta})$ and $h_{2}^* = h_2(\boldsymbol{\hat\uptheta})$ when computing $p(z)$.

Because there is no analytical solution for the integration of Eqn. (15), we compute the value of the integration by using Monte Carlo simulations and thus generate the pdf of a random route's travel time with two levels of uncertainties taken into consideration. The sketch of the algorithm for generating $p(z)$ is listed in Algorithm 1, where $N$ represents the total number of Monte Carlo simulations. The lower and upper bounds for $z$ are set as $a$ and $b$, and in real-world applications, $a$ and $b$ are the smallest and biggest values that we can choose to ensure $p(a)$ and $p(b)$ are sufficiently small. To facilitate the application of the $p(z)$ to construct travel time reliability measures, we propose to transfer $p(z)$ from the log(time) domain back into the time domain, and represent it in a discretized fashion. That is to say, by letting $t$ represent the pre-transformed travel time, i.e., $t = \exp(z)$, the final outputs of this algorithm are $[t_1, t_2, ..., t_K]$ and the probability at each $t_k$, denoted by $q(t_k), k = 1, 2, ..., K$. In Algorithm 1, after obtaining the probability density at $z$, i.e., $p(z)$, we show how to use it to compute the probability at the corresponding $t$. We are, therefore, simultaneously converting $z$ in the log(time) domain to $t$ in the time domain and converting from $p(z)$, the pdf for $z$, to $q(t), t = t_1, t_2, ..., t_K,$ the probability mass function (pmf) that closely approximates the probabilities that would be computed from the true pdf for the collection of all $t$ in a neighborhood of the selected discrete time points (the $t_k$).

\begin{algorithm}[ht]
\SetAlgoLined
\KwResult{$(t, q(t))$}
$z \gets a$\;
 \While{$z \leq b$}{
    $i \gets 1$\;
    \While{$i \leq N$}{
        Sample $h_{2i}$ from $N(h_2(\hat{\boldsymbol\uptheta}), [\boldsymbol{h}_2'(\hat{\boldsymbol\uptheta})]^T \boldsymbol{F}(\hat{\boldsymbol{\uptheta}})^{-1} \boldsymbol{h}_2'(\hat{\boldsymbol\uptheta}))$\;
        Compute $p(z|h_{2i})$\;
        $i \gets i+1$\;
    }
  $p(z) \gets \frac{1}{N}\sum_{i = 1}^{N} p(z|h_{2i})$\;
  $t = \exp(z)$\;
  $q(t) = p(z) \times \alpha $\;
  Output $(t, q(t))$\;
  $z \gets z + \upalpha$\;
 }
 \caption{Algorithm for generating pmf $q(t), t = t_1, t_2, ..., t_K$.}
\end{algorithm}

In Algorithm 1, $\upalpha$ indicates the step size of $z$ increment from $a$ to $b$, and $\upalpha$ is chosen by the users, which should be sufficiently small in order to capture $p(z)$ accurately. Notably, it may be better to apply a variable increment instead of a uniform one (like we did here), where more points should be sampled in rapidly changing regions. Importance sampling could also be used here \cite[Chap. 5]{rubinstein2016simulation}. After obtaining $q(t), t = t_1, t_2, ..., t_K$, the results can be directly used to construct different travel time reliability measures, which will be further discussed in the following subsection.


\subsection{Travel Time Reliability Measures}

We now have $q(t)$, the travel time probability distribution that captures the two-level uncertainties at each $t$ we sampled. Then, we apply $q(t)$ to compute various travel time reliability measures. There are multiple travel time reliability measures in the field (e.g., \cite{fha2010travel,pu2011analytic}), and in our paper, we use the 95th percentile travel times, standard deviation, coefficient of variation, buffer index, and planning time index to quantify travel time reliability.


For the travel time reliability measures, the 95th percentile travel times in $t$, say $t(0.95)$, indicating the probability of $t_k \leq t(0.95), k = 1, 2, ..., K$, is 0.95, i.e.,  
\begin{equation*}
    q(t_1) + q(t_2) + ... + q(t(0.95)) = 0.95.
\end{equation*}
The standard deviation, $\upsigma_t$, is defined as
\begin{equation*}
    \upsigma_t = \sqrt{E(t^2) - [E(t)]^2},
\end{equation*}
where $E(t)$ is the mean of $t$, i.e.,
\begin{equation*}
   E(t) = \sum_{k=1}^K t_k \cdot q(t_k),
\end{equation*}
and $E(t^2)$ is the second moment of $t$, i.e.,
\begin{equation*}
   E(t^2) = \sum_{k=1}^K t_k^2 \cdot q(t_k).
\end{equation*}
The coefficient of variation is defined as the ratio of the standard deviation $\upsigma_t$ to the sample mean $E(t)$, i.e.,
\begin{equation*}
    \text{Coefficient of variation} = \upsigma_t/E(t). 
\end{equation*}
The buffer index is defined as 
\begin{equation*}
    \text{Buffer index} = \frac{t(0.95) - E(t)}{E(t)}.
\end{equation*}
The planning time index is defined as the ratio of the 95th percentile travel times to the free-flow travel times, where we consider the 15th percentile travel times, say $t(0.15)$, as the free-flow travel time \cite{pu2011analytic}. That is
\begin{equation*}
    \text{Planning time index} = t(0.95)/t(0.15).
\end{equation*}

Generally speaking, the higher the travel time reliability is, the lower the standard deviation, coefficient of variation, buffer index, and planning time index are.

\section{Case Study}

Let us now give a case study for downtown Baltimore to illustrate the overall framework. We illustrate how it is possible to formally integrate route and link data in order to estimate the success probabilities of links and the associated (derived) network-wide parameters, and construct travel time reliability measures.

The selected transportation network is as shown in Fig. 2, where there are 46 links within the network. Note that the street between node I and node E and the street between node E and node A are one-way streets. So only two links (i.e., link IE and link EA) are considered for these two streets. The selected transportation network is located at downtown Baltimore, approximately 1 mile east of the center of the Inner Harbor area. This area is filled with local businesses and suffers traffic congestion problems, especially during rush hours.

\begin{figure}[H]
    \centering
    \includegraphics[width=0.48\textwidth]{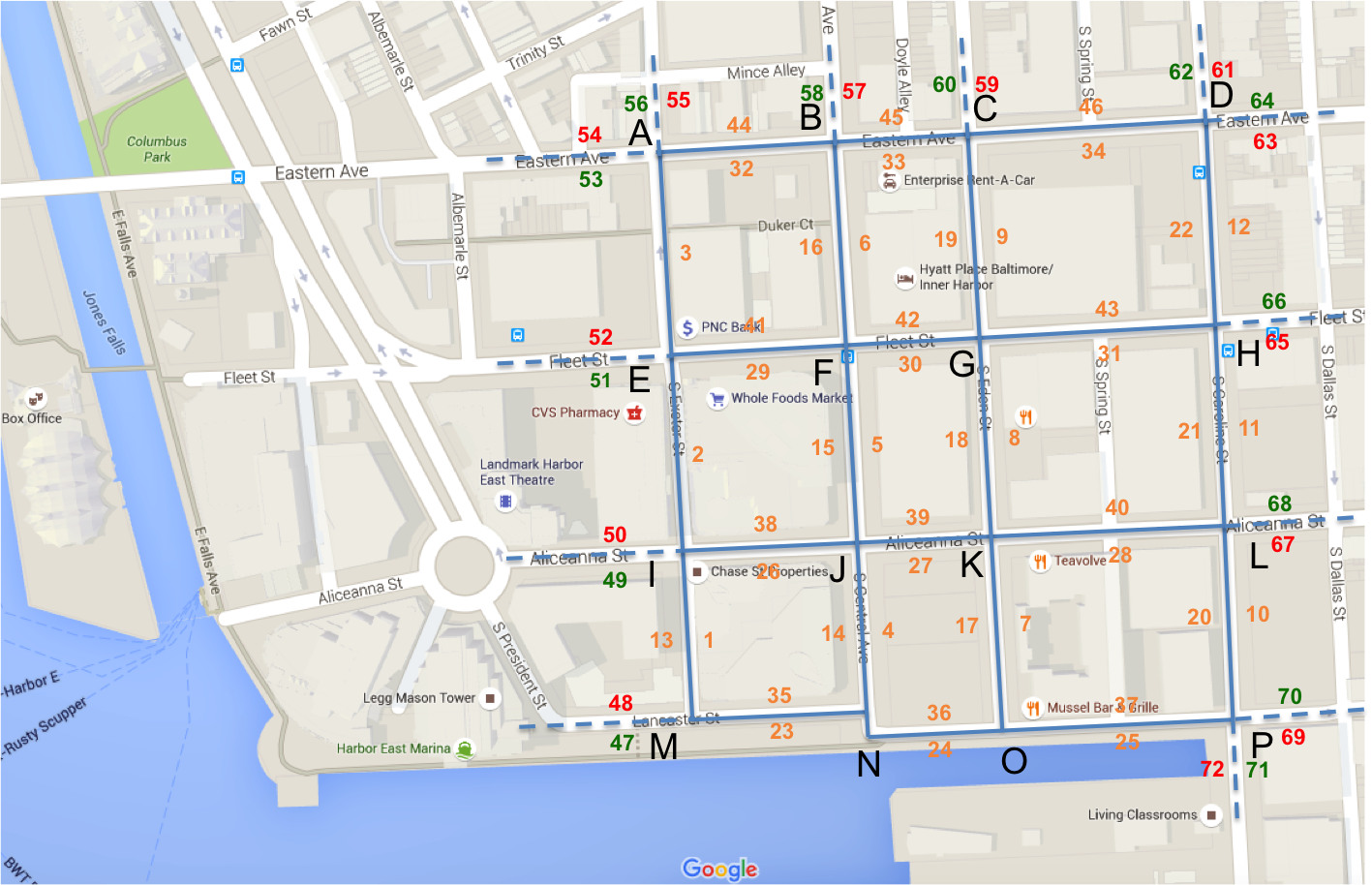}
    \caption{A transportation network in downtown Baltimore (approximately 1 mile east of the center of the Inner Harbor area): solid lines denote network of interest (A-B-C-D-E-F-G-H-I-J-K-L-M-N-O-P).}
    \label{fig:Bal}
\end{figure}

\subsection{Technical Approach Illustration}

As shown in Fig. 3, we use a simple flowchart to illustrate the overall technical approach introduced in Sects. II and III. There are three major steps in order to obtain the traffic condition estimates for links and travel time reliability for selected routes in the network. This subsection aims at providing a straightforward step-by-step guideline for transportation planners and engineers to directly use the methods proposed in this paper.

The first step is to collect data from Google Maps for the network. It is advisable to follow Subects. II-C and IV-B to collect link data (traffic conditions), and follow Subsect. II-D to collect route data (travel times). After properly collecting historical data from Google Maps, the second step is to compute MLEs that integrate route and link information by solving $\partial \log{L}/\partial \boldsymbol{\uptheta} = \boldsymbol{0}$ from Eqn. (4). The third step is to analyze travel time reliability for the route of interest by generating $q(t)$ using Algorithm 1 and then using $q(t)$ to construct different travel time reliability measures as described in Subsect. IV-D.

\begin{figure}[ht]
    \centering
    \includegraphics[width=0.38\textwidth]{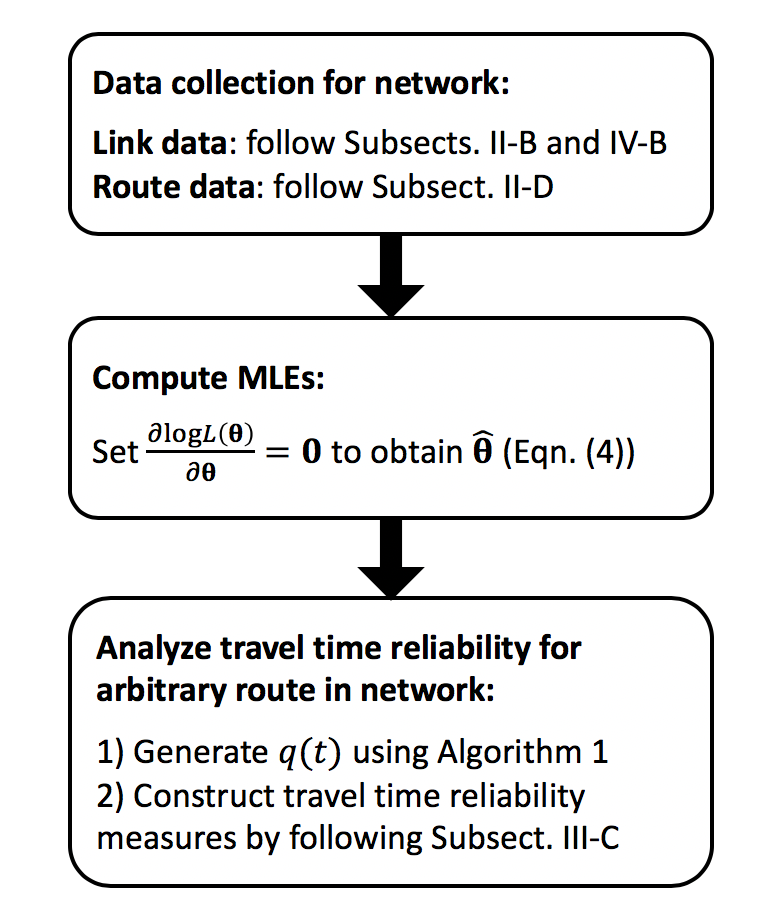}
    \caption{Flowchart of technical approach.}
    \label{fig:Bal}
\end{figure}

\subsection{Data Collection Strategy}

We collect route data on selected days and link data on other days in order to avoid implicit double counting of links. For the routes, we collect travel time data for different routes in the network on the same day (Google Maps allows for a choice of specific routes and then provides real-time travel time estimates based on the route choices). In this case study, we consider 12 routes in collecting data for this network (M-N-O-P-L-H-D; M-I-E-A-B-C-D; D-H-L-P-O-N-M; D-C-B-A; M-I-J-K-L-H; I-E-F-G-H-D; M-N-J-F-B-C; C-B-F-J-N-M; N-O-K-G-C-D; D-C-G-K-O-M; D-H-G-F-E; H-L-K-J-I-M). These 12 routes were chosen because the 12 routes cover all 46 traffic links within the network and because the 12 routes have few overlaps of links in order to minimize correlation among routes. Formal experimental design for how to choose routes to collect data might be beneficial for this study, but we do not consider such methods here. 

For the links, we collect real-time ``color'' data, representing live traffic conditions in four categories. That is, we use the Google Maps color scheme: green = normal traffic conditions, yellow = slower traffic conditions, red = congestion, and dark red = nearly stopped or stop-and-go traffic shown on links by choosing the ``traffic'' option in the menu of Google Maps. Here, we simplify the color scheme into binary states: a green or yellow is a ``1'' (success) while a red or dark red is a ``0'' (failure). When collecting the link data, we need to minimize the dependence among those links. Therefore, we propose the following link data collection strategy: by splitting the links within the network into two subsets (see Fig. 4), we collect data points for either Subset 1 or Subset 2 on one day, which guarantees data for Subset 1 are independent of data for Subset 2. Moreover, within a subset, there are no adjacent links considered; for instance, as shown in Fig. 4(a), link JK (considered in Subset 1) is directly connected to link KG, KL, and KO, none of which are considered in Subset 1. In this case, the link data collection strategy reduces the dependence of data between the two subsets; also, within a subset, links are not directly connected to each other, helping to reduce dependence.

\begin{figure}[!ht]  
\centering  
\begin{subfigure}{.4\textwidth}  
  \centering  
  \includegraphics[width=0.75\linewidth]{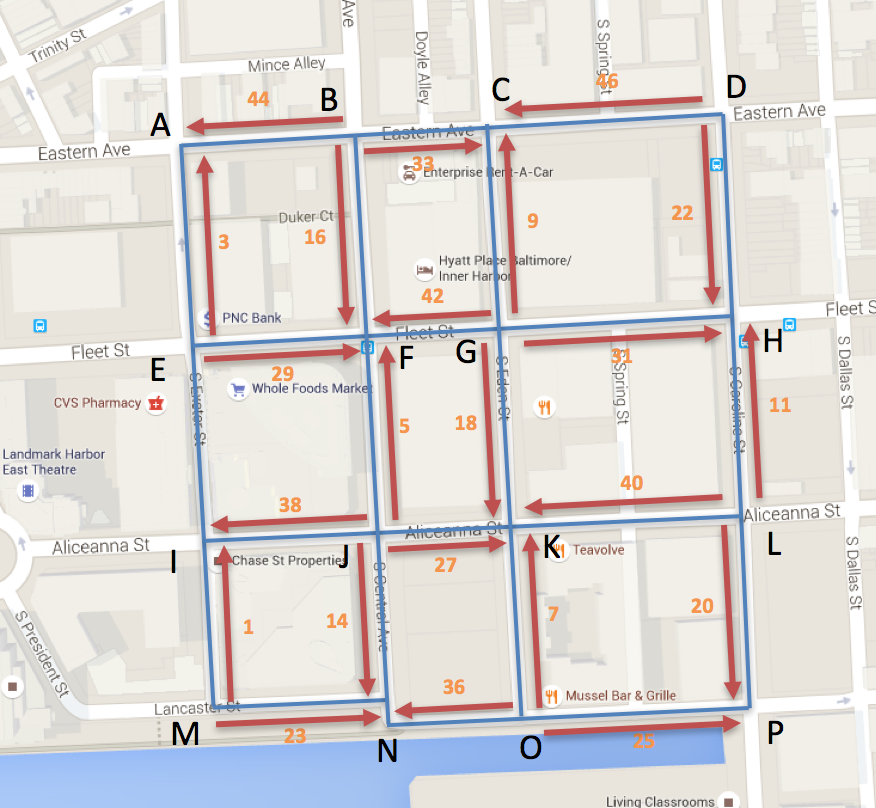}  
    \caption{Subset 1}   
\end{subfigure}   

\begin{subfigure}{.4\textwidth}  
  \centering  
  \includegraphics[width=0.75\linewidth]{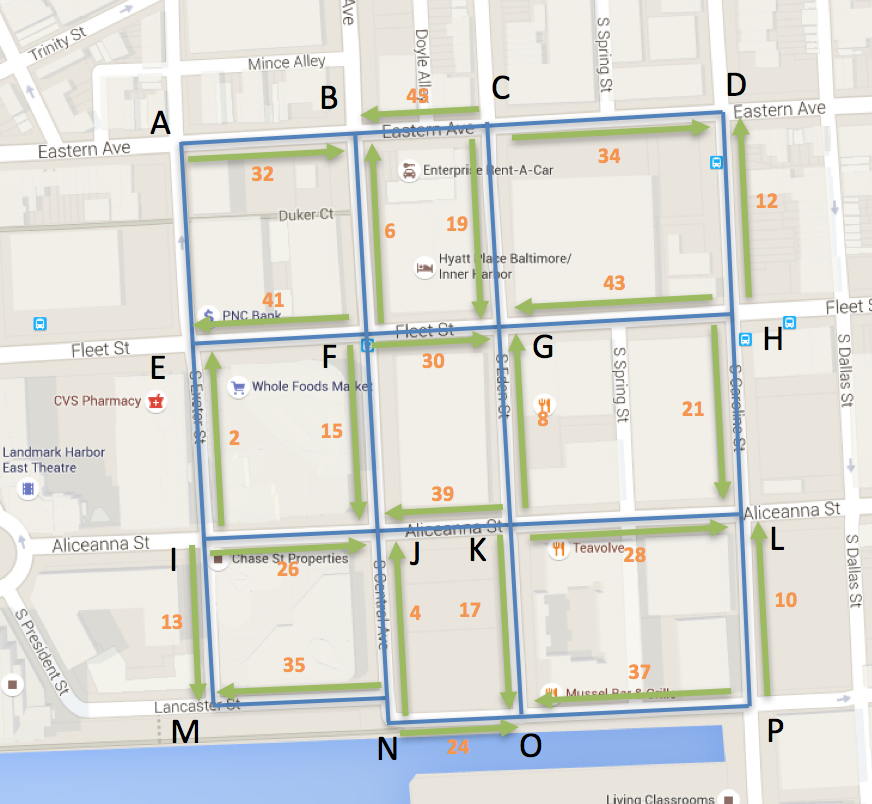}  
    \caption{Subset 2}   
\end{subfigure}  
\caption{Link data collection strategy: data for links in Subset 1 are collected on different days than links in Subset 2.}  
\end{figure}  

In this study, we collected 16 observations for each route and 11--27 observations for each link from Google Maps for this network at 5pm on certain weekdays (from Monday to Friday except for U.S. legal holidays) from March 31, 2016 through December 16, 2016. In order to evaluate the performance of this data collection strategy, we provide the following empirical evidence by applying hypothesis testing to test whether the link data are independent. 

In particular, we use Barnard's exact test \cite{Barnard1947} to test the null hypothesis that any link pairs within a subset (Subset 1 or 2) are independent versus the alternative that any link pairs within subset (Subset 1 or 2) are not independent. We compute the appropriate test statistic and associated $P$-values (probability values) for all 415 link pairs within each of the two subsets, and there are 15 out of 415 (around 4\%) that are below the 0.05 threshold. The 0.05 threshold indicates if the data are independent, we should expect 5\% of the test statistics to have $P$-values lower than 0.05. In this case, we have 4\% (quite close to 5\%) of the $P$-values lower than 0.05 threshold, which is consistent with the assumption of the independence of data. In addition to the overall assessment above, and in order to reduce the multiple comparison problem, we use the Bonferroni correction \cite{Rice2006} to test each individual hypothesis at $0.05/415 = 0.00012$. The smallest $P$-value obtained above is 0.0053 (much larger than 0.00012), so the Bonferroni correction cannot reject the null hypothesis. That is, by following the above data collection strategy, the empirical evidence is consistent with the hypothesis that link data are independent.

\subsection{MLE Results}

After properly collecting data for the network, we compute the MLEs for link success probabilities as shown in Appendix A. In contrast to the indicated sample means from only link data, it is expected that the MLEs for the links better represent the true success probabilities since the MLEs incorporate link interactions within the route data that may not be present in the color-coded link data. Taking link 46 as an example, the sample mean for link data alone is 1.00, but after incorporating route information, the MLE for the success probability in link 46 decreases to 0.80. This result is interesting, because for the days we collected the travel time data for routes, there existed some ``yellow'' traffic conditions for link 46. When modeling the link success/failure, we treated ``yellow'' link as a success; however, ``yellow'' represents slower traffic conditions that can be considered as a semi-failure. Only collecting the link data cannot capture the ``less-than-perfect'' traffic condition of link 46, but after integrating route and link data using MLE, we can model the link traffic condition more realistically.

As one application of the above, we are able to identify vulnerable links (low link success probabilities) within this network. For instance, in this network, link 2 has relatively low success probability, 0.54, compared to other links within the network. We may also notice that the success probability of link 2 is substantially lower than the links connected to it, namely, link 1, link 3, link 29, and link 38. Traffic engineers might wish to look into link 2 to figure out the reason for the vulnerability of this link in order to improve mobility.

\subsection{Travel Time Reliability Measures}

We select two routes (one with route data collected, the other without route data collected) to illustrate how to construct these travel time reliability measures in detail. To be specific, the two routes we pick are Route M-N-O-P-L-H-D (route data were collected), denoted as Route 1, and Route M-I-J-F-G (route data were not collected), Route 2. In this study, we use the travel time reliability measures previously defined in Subsect. III-C.

The results for travel time reliability measures are listed in Table 1. For the first route, we also compare the results obtained by our method with the results computed by using the empirical route data. Specifically, for Route 1 (M-N-O-P-L-H-D), the 95th percentile travel times computed by our method (4.41 min) is higher than the result obtained from the empirical data (4.00 min), while other measures obtained by our method are lower -- showing higher reliability of travel times for this specific route. In addition, after integrating route and link data, our method presents a more precise estimate with lower uncertainty.

Furthermore, for Route 2 (M-I-J-F-G), without collecting any empirical data for this route, our method produces a reasonable estimate. This result is very encouraging, since it is nearly impossible to collect empirical travel time data for all the routes in a large network. By using our method, one can estimate travel time and its corresponding reliability by collecting limited amount of link (traffic) and route (travel time) data. It is notable that our method has the potential to be applied to travel time reliability estimation for travel demand models (i.e., O-D level reliability) by integrating link and route data using MLE and developing a link data collection strategy to reduce the data dependence issue.

Another interesting observation is that even though Route 2 is shorter than Route 1, our results show that, compared with Route 1, the travel time reliability of Route 2 is lower: standard deviation, coefficient of variation, buffer index, and planning time index are all higher. One possible explanation is that, in Route 2, link 26 (IJ) is quite unreliable, having MLE for the link's success probability equal to 0.52; in contrast, all the links in Route 1 have high MLE values, ranging from 0.84 to 0.95.

\begin{table}[ht]
\footnotesize
\centering
\caption{Travel time reliability measures for two select routes: Route 1 (M-N-O-P-L-H-D): route data collected; Route 2 (M-I-J-F-G): no route data collected.}
\begin{tabular}{|c|c|c|c|}
\hline 
\multirow{2}{*}{Measures} & Route 1  & Route 1  & Route 2 \\ 
& (Route data) & (Our Method) & (Our Method) \\
\hline
95th percentile & \multirow{2}{*}{4.00 min} & \multirow{2}{*}{4.41 min} & \multirow{2}{*}{3.12 min}\\ 
travel times &  &&  \\
\hline
Standard & \multirow{2}{*}{0.48 min} & \multirow{2}{*}{0.34 min} & \multirow{2}{*}{0.35 min}\\
deviation &&&\\
\hline
Coefficient  & \multirow{2}{*}{0.14} & \multirow{2}{*}{0.09} & \multirow{2}{*}{0.14}\\ 
of variation &&&\\
\hline
Buffer index & 0.21 & 0.15 & 0.24\\ 
\hline
Planning time & \multirow{2}{*}{1.33} & \multirow{2}{*}{1.27} & \multirow{2}{*}{1.45}\\ 
index &&&\\
\hline
\end{tabular}
\end{table}

\section{Validation}

Due to small sample size of the case study presented in Sect. IV, in this section, we use the historical data from another network used in our previous study \cite{Zhao2016} to validate the MLE-based route/link approach. This dataset includes 54 data points for the route, and 18--19 data points for the links (after applying the link data collection strategy). We randomly split the route dataset into two subsets, with each one including 27 data points. One subset is used to estimate the MLEs using the proposed approach, and the other (independent) subset is used for testing. The results are shown in Table II, where the mean estimates are obtained using MLE (our method) and sample mean (testing data), and the standard deviation of MLE is measured by $\sqrt{\boldsymbol{\upxi}'(\boldsymbol{\hat\uptheta})^T \boldsymbol{F}(\boldsymbol{\hat\uptheta})^{-1} \boldsymbol{\upxi}'(\boldsymbol{\hat\uptheta})}$ (where $\upxi(\boldsymbol{\uptheta}) = \sum_{j=1}^{m(q)}\big[l_{q_j}/v'-l_{q_j}\uprho_{q_j}(v-v')/(vv')\big]$) and the standard deviation of sample mean is measured by the sample standard deviation over the square root of testing data sample size.

Table II shows that our method (10.30 min) produces a more accurate estimate for the testing data sample mean (10.15 min), compared to the training data sample mean (10.48 min). The MLE is included within the one standard deviation interval for the testing data sample mean (9.97 min--10.33 min), while the training data sample mean is barely covered by the two standard deviations interval for the testing data sample mean (9.79 min--10.51 min). Moreover, the standard deviation of the MLE (0.10 min) is much lower than the training data's standard deviation (0.20 min), showing MLE produces a more precise estimate. Hence, by integrating route and link data using MLE, our method produces a more accurate and precise estimation compared to the sample mean of the empirical data.

\begin{table}[H]
\small
\centering
\caption{Validation results.}
\begin{tabular}{|c|c|c|}
\hline
& \multirow{2}{*}{Mean estimate}   & Standard deviation   \\ 
& & of mean estimate \\ \hline
Our method     & \multirow{2}{*}{10.30 min}    & \multirow{2}{*}{0.10 min}       \\
(MLE) & & \\ \hline
Training data  & \multirow{2}{*}{10.48 min}   & \multirow{2}{*}{0.20 min}       \\ 
(Sample mean) && \\ \hline
Testing data    & \multirow{2}{*}{10.15 min}     & \multirow{2}{*}{0.18 min}       \\
(Sample mean) && \\ \hline
\end{tabular}
\end{table}

\section{Conclusion}

Above all, we propose a novel method to model transportation networks by using statistical methods to integrate route and link data collected from real-time navigation services (such as Google Maps). Then, we use the properties of our model to generate the probability distribution of travel times for an arbitrary route within the network, and apply the distribution to construct travel time reliability measures for this route. This paper provides a tool for practitioners to better model traffic dynamics \cite{Zhao2018} and forecast travel demand for potential use in planning policy interventions or helping reduce traffic congestion problems. 

A notable limitation of this paper is that route data and link data are assumed to be independent. Of course, independence is also an issue in other methods such as \cite{ma2017estimation} and \cite{gupta2018incorporation}. In practice, it is possible to separate links and routes spatially and across days to ensure that the independence assumption is at least approximately valid, like we did in this paper. Furthermore, the case study provided in this paper is relatively small. In future work, we plan to resolve these two major limitations by conducting experimental design for data collection to ensure independence and apply this model to a bigger network in the real world. We also want to convert this model into a user-friendly decision-support tool in order to support more practitioners and help advance the state-of-the-art of traffic planning.


%

\appendices
\section{Estimation Results for Network in Downtown Baltimore}

The sample means below are the estimates of $\uprho_j$ from data on link $j$ only; the MLEs are the estimates from link and route data. The relative difference shows the percentage difference relative to the sample mean.

\begin{table}[H]
\footnotesize
\centering
\begin{tabular}{||c|c|c|c||}
\hline 
Link No. & Sample Mean & MLE & Relative Difference \\ 
\hline
\hline
1 & 0.80	& 0.77 &	$-4.4$\% \\ 
\hline 
2 & 0.50	& 0.54	& 7.3\%  \\ 
\hline 
3 & 0.78 & 0.78 & $-0.0$\%  \\ 
\hline 
4 & 0.65	& 0.61	& $-6.3$\%  \\ 
\hline 
5 & 0.89	& 0.90	& 1.1\%  \\ 
\hline 
6 & 0.88	& 0.89	& 1.2\%  \\ 
\hline 
7 & 0.63	& 0.68	& 9.1\%  \\ 
\hline 
8 & 0.64	& 0.68	& 5.8\%  \\ 
\hline 
9 & 0.56	& 0.60	& 8.3\%  \\ 
\hline 
10 & 0.88 & 0.88 & $-1.1$\%  \\ 
\hline 
11 & 0.89 & 0.84 & $-5.2$\%  \\ 
\hline 
12 & 0.96 & 0.91 & $-5.1$\%  \\ 
\hline 
13 & 0.55 & 0.56 & 2.6\%  \\ 
\hline 
14 & 1.00 & 0.98 & $-1.6$\%  \\ 
\hline 
15 & 0.85 & 0.82 & $-2.6$\%  \\ 
\hline 
16 & 0.74 & 0.76 & 2.8\%  \\ 
\hline 
17 & 0.69 & 0.84 & 22.0\%  \\ 
\hline 
18 & 0.70 & 0.82 & 16.5\%  \\ 
\hline 
19 & 0.64 & 0.76 & 20.1\%  \\ 
\hline 
20 & 0.96 & 0.95 & $-1.1$\%  \\ 
\hline 
21 & 0.77 & 0.82 & 6.0\%  \\ 
\hline 
22 & 0.70 & 0.82 & 16.9\%  \\ 
\hline 
23 & 0.89 & 0.89 & 0.5\%  \\ 
\hline 
24 & 0.92 & 0.93 & 0.9\% \\ 
\hline 
25 & 1.00 &	0.95 &	$-4.7$\% \\ 
\hline 
26 & 0.50	& 0.52	& 4.2\% \\ 
\hline 
27 & 0.89	& 0.87	& $-1.9$\% \\ 
\hline 
28 & 0.42 & 0.47	& 11.6\% \\ 
\hline 
29 & 0.85	& 0.84	& $-1.9$\% \\ 
\hline 
30 & 0.81 & 0.80 & $-0.5$\% \\ 
\hline 
31 & 0.81 & 0.78 & $-4.3$\% \\ 
\hline 
32 & 0.96 & 0.96 & 0.0\% \\ 
\hline 
33 & 1.00 & 1.00	& 0.0\% \\ 
\hline 
34 & 0.96	& 0.96	& 0.1\% \\ 
\hline 
35 & 0.96	& 0.94	& $-2.4$\% \\ 
\hline 
36 & 0.96	& 0.98	& 1.4\% \\ 
\hline 
37 & 0.88	& 0.86	& $-2.5$\% \\ 
\hline 
38 & 0.96	& 0.95	& $-1.0$\% \\ 
\hline 
39 & 0.88 & 0.88 & $-1.0$\% \\ 
\hline 
40 & 0.96 & 0.94 & $-2.1$\% \\ 
\hline 
41 & 0.81 & 0.85 & 5.8\% \\ 
\hline 
42 & 0.67 & 0.77 & 15.0\% \\ 
\hline 
43 & 0.92	& 0.91	& $-1.3$\% \\ 
\hline 
44 & 0.85	& 0.81	& $-4.8$\% \\ 
\hline 
45 & 0.81	& 0.86	& 6.5\% \\ 
\hline 
46 & 1.00	& 0.80	& $-20.2$\% \\ 
\hline 
\end{tabular} 
\end{table}


\section*{Acknowledgment}

The authors would like to acknowledge the financial support from the Johns Hopkins University Applied Physics Laboratory IRAD Program and the National Science Foundation Grant RIPS 1441209. Any opinions, findings, and conclusions or recommendations expressed in this material are those of the authors and do not necessarily reflect the views of the funding organizations. We also thank Xiang Yan for providing useful suggestions from a planner's perspective.

\ifCLASSOPTIONcaptionsoff
  \newpage
\fi



\bibliographystyle{IEEEtran}
\bibliography{references}

\begin{thebibliography}{10}
\providecommand{\url}[1]{#1}
\csname url@samestyle\endcsname
\providecommand{\newblock}{\relax}
\providecommand{\bibinfo}[2]{#2}
\providecommand{\BIBentrySTDinterwordspacing}{\spaceskip=0pt\relax}
\providecommand{\BIBentryALTinterwordstretchfactor}{4}
\providecommand{\BIBentryALTinterwordspacing}{\spaceskip=\fontdimen2\font plus
\BIBentryALTinterwordstretchfactor\fontdimen3\font minus
  \fontdimen4\font\relax}
\providecommand{\BIBforeignlanguage}[2]{{%
\expandafter\ifx\csname l@#1\endcsname\relax
\typeout{** WARNING: IEEEtran.bst: No hyphenation pattern has been}%
\typeout{** loaded for the language `#1'. Using the pattern for}%
\typeout{** the default language instead.}%
\else
\language=\csname l@#1\endcsname
\fi
#2}}
\providecommand{\BIBdecl}{\relax}
\BIBdecl

\bibitem{Vasserman2015}
S.~Vasserman, M.~Feldman, and A.~Hassidim, ``{Implementing the wisdom of
  \text{Waze}},'' in \emph{Proc. of The 24th International Joint Conference on
  Artificial Intelligence (IJCAI 2015)}, 2015, pp. 660--666.

\bibitem{Jeske2013}
T.~Jeske, ``{Floating car data from smartphones: What Google and Waze know
  about you and how hackers can control traffic},'' in \emph{Proc. of The
  BlackHat Europe}, 2013, pp. 1--12.

\bibitem{GoogleMaps}
\BIBentryALTinterwordspacing
{Google Maps}, ``{Google Maps APIs}.'' [Online]. Available:
  \url{https://enterprise.google.com/maps/products/mapsapi.html}
\BIBentrySTDinterwordspacing

\bibitem{Merchant1978}
\BIBentryALTinterwordspacing
D.~K. Merchant and G.~L. Nemhauser, ``{A model and an algorithm for the dynamic
  traffic assignment problems},'' \emph{Transportation Science}, vol.~12,
  no.~3, pp. 183--199, 1978. [Online]. Available:
  \url{http://transci.journal.informs.org/content/12/3/183.abstract}
\BIBentrySTDinterwordspacing

\bibitem{Daganzo1994}
C.~F. Daganzo, ``{The cell transmission model: A dynamic representation of
  highway traffic consistent with the hydrodynamic theory},''
  \emph{Transportation Research Part B: Methodological}, vol.~28, no.~4, pp.
  269--287, 1994.

\bibitem{kotsialos2002traffic}
A.~Kotsialos, M.~Papageorgiou, C.~Diakaki, Y.~Pavlis, and F.~Middelham,
  ``Traffic flow modeling of large-scale motorway networks using the
  macroscopic modeling tool \text{METANET},'' \emph{IEEE Transactions on
  Intelligent Transportation Systems}, vol.~3, no.~4, pp. 282--292, 2002.

\bibitem{celikoglu2007dynamic}
H.~B. Celikoglu, ``A dynamic network loading model for traffic dynamics
  modeling,'' \emph{IEEE Transactions on Intelligent Transportation Systems},
  vol.~8, no.~4, pp. 575--583, 2007.

\bibitem{Ben-Akiva2012}
\BIBentryALTinterwordspacing
M.~E. Ben-Akiva, S.~Gao, Z.~Wei, and Y.~Wen, ``{A dynamic traffic assignment
  model for highly congested urban networks},'' \emph{Transportation Research
  Part C: Emerging Technologies}, vol.~24, pp. 62--82, 2012. [Online].
  Available: \url{http://dx.doi.org/10.1016/j.trc.2012.02.006}
\BIBentrySTDinterwordspacing

\bibitem{ran2012dynamic}
B.~Ran and D.~Boyce, \emph{Dynamic Urban Transportation Network Models: Theory
  and Implications for Intelligent Vehicle-Highway Systems}.\hskip 1em plus
  0.5em minus 0.4em\relax Springer Science \& Business Media, 2012, vol. 417.

\bibitem{chiou2017robust}
S.-W. Chiou, ``Robust stochastic design of signal-controlled road network under
  uncertain travel demands,'' \emph{IEEE Transactions on Automatic Control},
  vol.~62, no.~7, pp. 3152--3164, 2017.

\bibitem{du2018traffic}
L.~Du, G.~Song, Y.~Wang, J.~Huang, M.~Ruan, and Z.~Yu, ``Traffic events
  oriented dynamic traffic assignment model for expressway network: A network
  flow approach,'' \emph{IEEE Intelligent Transportation Systems Magazine},
  vol.~10, no.~1, pp. 107--120, 2018.

\bibitem{Spall2014}
J.~C. Spall, ``{Identification for systems with binary subsystems},''
  \emph{IEEE Transactions on Automatic Control}, vol.~59, no.~1, pp. 3--17,
  2014.

\bibitem{Spall2012}
------, ``{Asymptotic Normality and Uncertainty Bounds for Reliability
  Estimates from Subsystem and Full System Tests},'' in \emph{Proc. of American
  Control Conference}.\hskip 1em plus 0.5em minus 0.4em\relax Montreal, Canada:
  IEEE, 2012, pp. 56--61.

\bibitem{Spall2013}
------, ``{Parameter estimation for systems with binary subsystems},'' in
  \emph{Proc. of American Control Conference}.\hskip 1em plus 0.5em minus
  0.4em\relax Washington D.C.: IEEE, 2013, pp. 83--88.

\bibitem{Zhao2016}
X.~Zhao and J.~C. Spall, ``{Estimating travel time in urban traffic by modeling
  transportation network systems with binary subsystems},'' in \emph{Proc. of
  American Control Conference}, 2016, pp. 803--808.

\bibitem{fha2010travel}
U.~S. D. o.~T. Federal Highway~Administration, ``Travel time reliability:
  Making it there on time, all the time,'' 2006.

\bibitem{carrion2012value}
C.~Carrion and D.~Levinson, ``Value of travel time reliability: A review of
  current evidence,'' \emph{Transportation Research Part A: Policy and
  Practice}, vol.~46, no.~4, pp. 720--741, 2012.

\bibitem{chen2003travel}
C.~Chen, A.~Skabardonis, and P.~Varaiya, ``Travel-time reliability as a measure
  of service,'' \emph{Transportation Research Record: Journal of the
  Transportation Research Board}, no. 1855, pp. 74--79, 2003.

\bibitem{clark2005modelling}
S.~Clark and D.~Watling, ``Modelling network travel time reliability under
  stochastic demand,'' \emph{Transportation Research Part B: Methodological},
  vol.~39, no.~2, pp. 119--140, 2005.

\bibitem{pu2011analytic}
W.~Pu, ``Analytic relationships between travel time reliability measures,''
  \emph{Transportation Research Record: Journal of the Transportation Research
  Board}, no. 2254, pp. 122--130, 2011.

\bibitem{uchida2015travel}
K.~Uchida, ``Travel time reliability estimation model using observed link flows
  in a road network.'' \emph{Comp.-Aided Civil and Infrastruct. Engineering},
  vol.~30, no.~6, pp. 449--463, 2015.

\bibitem{gupta2018incorporation}
S.~Gupta, P.~Vovsha, A.~Dutta, V.~Livshits, W.~Zhang, and H.~Zhu,
  ``Incorporation of travel time reliability in regional travel model,''
  \emph{Transportation Research Record: Journal of the Transportation Research
  Board}, 2018.

\bibitem{Scholz2006}
F.~W. Scholz, ``{Maximum likelihood estimation},'' in \emph{Encyclopedia of
  Statistical Sciences, 7}.\hskip 1em plus 0.5em minus 0.4em\relax John Wiley
  {\&} Sons, Inc., 2006.

\bibitem{Rice2006}
J.~A. Rice, \emph{{Mathematical Statistics and Data Analysis}}, 3rd~ed.\hskip
  1em plus 0.5em minus 0.4em\relax Nelson Education, 2006.

\bibitem{ElFaouzi2007}
N.~E. {El Faouzi} and M.~Maurin, ``{Reliability metrics for path travel time
  under log-normal distribution},'' in \emph{Proc. of The 3rd International
  Symposium on Transportation Network Reliability}, 2007.

\bibitem{Spall2003}
J.~C. Spall, \emph{{Introduction to Stochastic Search and Optimization:
  Estimation, Simulation, and Control}}.\hskip 1em plus 0.5em minus 0.4em\relax
  Wiley, 2003.

\bibitem{Johnson1994}
N.~L. Johnson, S.~Kotz, and N.~Balakrishnan, \emph{{Continuous Univariate
  Distributions, Vol. 1}}.\hskip 1em plus 0.5em minus 0.4em\relax New York:
  Wiley, 1994.

\bibitem{rubinstein2016simulation}
R.~Y. Rubinstein and D.~P. Kroese, \emph{Simulation and the Monte Carlo
  Method}.\hskip 1em plus 0.5em minus 0.4em\relax John Wiley \& Sons, 2017.

\bibitem{Barnard1947}
G.~A. Barnard, ``{Significance tests for 2x2 tables},'' \emph{Biometrika},
  vol.~34, pp. 123--138, 1947.

\bibitem{Zhao2018}
X.~Zhao and J.~C. Spall, ``A \text{Markovian} framework for modeling dynamic
  network traffic,'' in \emph{Proc. of Annual American Control
  Conference}.\hskip 1em plus 0.5em minus 0.4em\relax Milwaukee, WI: IEEE,
  2018, pp. 6616--6621.

\bibitem{ma2017estimation}
Z.~Ma, H.~N. Koutsopoulos, L.~Ferreira, and M.~Mesbah, ``Estimation of trip
  travel time distribution using a generalized \text{Markov} chain approach,''
  \emph{Transportation Research Part C: Emerging Technologies}, vol.~74, pp.
  1--21, 2017.

\end{thebibliography}
%

%

\end{document}